\documentclass[12pt]{article}
\usepackage{amsmath}
\usepackage{amssymb}
\usepackage{epsfig}
\usepackage{euscript}
\usepackage{fancybox}
\usepackage{color}
\def\mathswitchr#1{\relax\ifmmode{\mathrm{#1}}\else$\mathrm{#1}$\fi}

\newcommand {\pslash}{\hbox{$\not\hbox{\kern-2.3pt $p$}$}}

\usepackage{cite}

\usepackage{epic}
\def\alf1{ {\alpha\over\pi} }
\begin{document}
%\input{feynman} 
%=======================================================================
\begin{titlepage}
\begin{flushright}
%{\bf MPI-PhT-2002-08}\\
 {\bf BU-HEPP-03-11,UTHEP-03-1101 }\\
{\bf Nov, 2003}\\
\end{flushright}
%\vspace{0.05cm}
 
\begin{center}
{\Large Massive Elementary Particles and Black Holes$^{\dagger}$
}
\end{center}

\vspace{2mm}
\begin{center}
%%  {\bf   S. Jadach$^{a,b}$ and B.F.L. Ward$^{c,d}$}
{\bf   B.F.L. Ward}\\
\vspace{2mm}
%{\em $^a$CERN, Theory Division, CH-1211 Geneva 23, Switzerland,}\\
%{\em $^b$Institute of Nuclear Physics,
%        ul. Kawiory 26a, Krak\'ow, Poland,}
%{\em $^c$Werner-Heisenberg-Institut, Max-Planck-Institut fuer Physik,
%Muenchen, Germany,}\\
%{\em $^a$Werner-Heisenberg-Institut, Max-Planck-Institut fuer Physik,
%Muenchen, Germany,}\\
%{\em $^d$Department of Physics and Astronomy,\\
%  The University of Tennessee, Knoxville, Tennessee 37996-1200, USA.}\\
%{\em $^c$SLAC, Stanford University, Stanford, California 94309, USA,}\\
{\em Department of Physics,\\
  Baylor University, Waco, Texas 76798-7316, USA}\\
{\em and }\\
{\em Department of Physics and Astronomy,\\
  The University of Tennessee, Knoxville, Tennessee 37996-1200, USA}\\
%{\em $^c$SLAC, Stanford University, Stanford, California 94309, USA,}\\
\end{center}

%\begin{center}
% {\bf S. Jadach}\\
%{DESY, Theory Division, D-22603 Hamburg, Germany}\\
%   {\em Institute of Nuclear Physics,
%        ul. Kawiory 26a, Krak\'ow, Poland}\\
%   {\em CERN, Theory Division, CH-1211 Geneva 23, Switzerland,}\\
% {\bf W.  P\l{a}czek}\\
%   {\em Institute of Computer Science,
%   Jagellonian University, ul. Nawojki 11, 30-072 Krak\'ow, Poland}\\
%{\em CERN, Theory Division, CH-1211 Geneva 23, Switzerland,}\\
% {\bf M. Skrzypek}\\
%   {\em Institute of Nuclear Physics,
%        ul. Kawiory 26a, Krak\'ow, Poland}\\
%{\em CERN, Theory Division, CH-1211 Geneva 23, Switzerland,}\\
% {\bf B.F.L. Ward}\\
%   {\em Department of Physics and Astronomy,\\
%   The University of Tennessee, Knoxville, Tennessee 37996-1200\\
%   SLAC, Stanford University, Stanford, California 94309}\\
%{\em CERN, Theory Division, CH-1211 Geneva 23, Switzerland,}\\
%{\bf Z. W\c as}\\
%   {\em Institute of Nuclear Physics,
%        ul. Kawiory 26a, Krak\'ow, Poland}\\
%   {\em CERN, Theory Division, CH-1211 Geneva 23, Switzerland}\\
%\end{center}

\vspace{5mm}
\begin{center}
{\bf   Abstract}
\end{center}
An outstanding problem posed by Einstein's general theory of relativity
to the quantum theory of point particle fields is the fate of a
massive point particle; for, in the classical solutions of
Einstein's theory, such a system should be a black hole.
We use exact results in a new approach to quantum gravity to 
show that this conclusion is obviated by quantum loop effects.
Phenomenological implications are discussed.\\
\vspace{10mm}
% 
%\vspace{10mm}
\\
\centerline{ Submitted to JCAP }
\renewcommand{\baselinestretch}{0.1}
\footnoterule
\noindent
{\footnotesize
\begin{itemize}
\item[${\dagger}$]
Work partly supported 
% the Polish Government
%grants KBN 2P30225206 and 2P03B17210, the Maria Sk\l{}odowska-Curie
%Joint Fund II PAA/DOE-97-316, and
by the US Department of Energy Contract  DE-FG05-91ER40627
and by NATO Grant PST.CLG.977751.
%, and by
%Polish Government grant 5P03B09320.
\end{itemize}
}
%\vspace{0.5cm}
%\centerline{ Submitted to JCAP }
%\begin{flushleft}
%{\bf UTHEP-00-0101}\\
%{\bf Jan, 2000}\\
%\end{flushleft}

\end{titlepage}

%=======================================================================
\def\Kmax{K_{\rm max}}\def\ieps{{i\epsilon}}\def\rQCD{{\rm QCD}}
\renewcommand{\theequation}{\arabic{equation}}
\font\fortssbx=cmssbx10 scaled \magstep2
\renewcommand\thepage{}
%\vfill\eject
\parskip.1truein\parindent=20pt\pagenumbering{arabic}\par
%\section{\bf Introduction}\label{intro}\par
The ideas of Albert Einstein in the special theory of relativity
have been incorporated into quantum mechanics so that we can say
the union of Niels Bohr and Albert Einstein has been achieved in the
theory of special relativistic point particle quantum fields.
Indeed, the proto-typical example of such a theory is 
the Standard Model ( SM )~\cite{sm,qcd1}
and its successes are a true triumph of the 20th century.
However, for Einstein's 
theory of general relativity, the situation regarding
its union with quantum mechanics is markedly different;
for, even though his general theory of relativity
has had also many successes~\cite{mtw,sw1},
it has so far evaded a complete, direct application 
of quantum mechanics. All of the accepted treatments
of the complete quantum loop corrections to Einstein's theory involve
recourse to as yet phenomenologically unfounded
theoretical paradigms~\cite{gsw,jp} which seem to imply even the modification
of quantum mechanics itself.
In Ref.~\cite{bw1}, we have introduced a new approach to
quantum gravity wherein we use resummation of large 
higher order radiative corrections to ameliorate
the apparently bad UV behavior of the 
theory. This new approach,
which relies on phenomenologically well-founded theoretical
paradigms and which does not involve any modification of
quantum mechanics itself, allows us 
to analyze truly complete finite quantum loop effects in
Einstein's general theory of relativity. In this paper, 
we present such an analysis.
\par
It is important to put our approach in the proper context
of the current literature on quantum general relativity,
which is an extensive literature indeed~\cite{qgenrel}.
The situation was already summarized by Weinberg in Ref.~\cite{wein1}.
There, Weinberg argued that there are basically four approaches to
the bad UV behavior of quantum general relativity:
\begin{itemize}
\item extended theories of gravitation: supersymmetric theories - superstrings
\item resummation 
\item composite gravitons
\item asymptotic safety: fixed point theory ( see Ref.~\cite{laut} )
%%%lautscher \& reuter,  hep-th/0205062)} 
\end{itemize}
In what follows, we discuss a new version of the resummation approach
based on the methods of Yennie, Frautschi and Suura (YFS) in 
Ref.~\cite{yfs}.
Since these methods are not generally familiar, we review their essence
as well in what follows.\par

The key element in the YFS approach to resummation is the rearrangement
of the respective series in such away that the large real and 
virtual corrections from the infrared regime are summed up to all orders.
In Abelian gauge theories, the resummation results in exponential factors
which contain all of the IR singularities and these exactly
cancel between real and virtual corrections in the exponential factors for
physical observables. In non-Abelian gauge theories, not all IR 
singularities exponentiate in the YFS expansion so that, after the
YFS rearrangement, the respective residuals still exhibit 
a final cancellation~\cite{qcdyfs} between real and virtual corrections
in the physical observables order by order in the loop expansion.
In the quantum theory of general relativity, wherein the symmetry group
is non-Abelian, application of the YFS methods then allows us to isolate
the large IR contributions to the loop expansion which do
exponentiate and resum them in an exact rearrangement of the respective
loop expansion~\cite{bw1}. The remarkable result is that, having done this,
the remaining expansion becomes much more 
convergent in the ultra-violet (UV)
regime than was the original expansion. Indeed, the UV infinities
are completely removed by the YFS resummation. It is this result
that we will apply in the discussion which follows.\par

It may seem somewhat paradoxical that the infrared regime of a theory
could produce, upon resummation, an exact rearrangement of the respective
perturbation series in which the ultra-violet regime
of the resummed series is much better behaved than the original
series. What one needs to recall here is that, in the loop expansion,
the integration over the 4-dimensional loop momentum, $\int d^4k$,
generates enhanced contributions in point particle quantum field theory
in three regimes:
\begin{itemize}
\item the infrared regime (IR)
\item the collinear regime (CL)
\item the ultra-violet regime (UV).
\end{itemize}
The usual renormalization program subtracts the UV divergent parts
of the integration and thereby renders the integral convergent in the
UV regime were it not already so. The remaining large contributions from the
IR and CL regimes then also make an enhanced
contribution to the result of the integration 
whenever they are divergent or posses would-be mass singularities
that would render them divergent when a mass goes to zero.
The latter result in big logs, $L\equiv \ln Q^2/m^2$, 
$\ell\equiv \ln Q^2/\lambda^2$, appearing to
an attendant power as a result of the integration, where we treat $Q$ as a
generic hard scale in the problem, m a typical external light particle
mass, and $\lambda$ as an infrared regulator mass for massless 
non-zero integer spinning particles. 
These big logs are just as large in general
as any big logs left-over from the UV subtraction in the 
renormalization program and hence are just as important, if not more
important, than the latter.
It is the YFS rearrangement of these important
IR terms in the virtual corrections that the new approach to quantum gravity
in Ref.~\cite{bw1} uses to tame the UV behavior of quantum general
relativity. It follows that, as the graviton couples to all point particles,
this type of improvement of the UV behavior found in Ref.~\cite{bw1} applies
to all renormalizable point particle quantum field theories and to a 
large class of non-renormalizable point particle quantum field theories
that includes quantum general relativity, of course.\par

Here, we need to stress that, while we are making a YFS resummation of the
theory of quantum general relativity, we are in fact focusing on the
UV regime of the theory, the regime of sub-Planck length physics.
This should be compared with the rather complementary efforts in Refs.
~\cite{dono1}, which use the analog of the chiral perturbation
theory familiar from the strong interaction and address then 
large distance limit of quantum general relativity. We see no contradiction
between the new resummed approach in Ref.~\cite{bw1} and the large
distance, effective Lagrangian methods of Refs.~\cite{dono1}. The
relationship between the two is the same as the relationship between
the use of perturbative QCD ( Quantum Chromodynamics~\cite{qcd1} )
for the hard strong interactions
and the use of chiral perturbation theory for the soft, large distance
strong interactions, where there should be then a matching between 
the respective large distance and short distance formalisms 
at some appropriate intermediate scale -- sometimes in 
QCD this has to be done
using fully non-perturbative methods and we see no reason why the
similar situation might not hold in the
quantum theory of general relativity. We hope to come back to these types
of analyses elsewhere~\cite{elswh}.\par 

Currently, the only {\it really accepted} 
approach to quantum general relativity
is that embodied in the well-known but incompletely understood superstring
theory~\cite{gsw,jp}. Therefore, one of the main results of the resummed
quantum gravity theory of Ref.~\cite{bw1}, using the 
superstring as a benchmark, is to provide an
alternative view of the important problem of quantum general relativity.
There is one fundamental difference between the two approaches:
in the superstring theory, the Planck length is special, and one really can not
readily consider physics below it whereas in the new, resummed quantum gravity
there is no problem at all to discuss sub-Planck scale physics. The situation
reminds one of the old string theory~\cite{schw1}
approach to the strong interaction in the 1960's and early 1970's, wherein
the size of a hadron, $\sim 1$fm, was special and it was difficult
to discuss sub-fermi physics. We of course have since learned that the
old string theory of the strong interaction was just a phenomenological
model for the point particle field theory of QCD, the truly fundamental
description of the strong interaction. Evidently, our new, resummed
quantum gravity theory suggests that the same story may occur again --
namely, the extended object theory, the superstring in this case, 
is really just a phenomenological model for a more
fundamental point particle quantum field theory that describes 
quantum general relativity to 
sub-Planck scale distances as well. Presumably, the superstring theory
can help us determine just what this truly fundamental theory,
which would describe all the known forces to arbitrary distances below
the Planck scale, would be. Here, we refer to it as {\bf TUT}, the ultimate
theory. We hope to participate in its construction elsewhere~\cite{elswh}.\par

At this point, the reader may be wondering how, if ever, observations
could probe the sub-Planck scale regime? We do know that quantum
loop corrections do probe arbitrary scales so that sufficiently precise
measurements of appropriately chosen quantities must be sensitive to
sub-Planck scale physics. It is also natural to speculate that the
the early universe may provide such observables? Here, however, there is
controversy. In Ref.~\cite{kolb}, it is stated that there are about
8 orders of magnitude between what scale astrophysical observations
can easily probe today and the Planck scale. This is not 
an insurmountable difference in scales, as either appropriate
observables and/or appropriate precision can easily
overcome as many as 16 orders of magnitude, as in proton decay
studies~\cite{proton}. We do, however, need to find other 
arenas in which to try to
develop tests of the physics in the sub-Planck scale regime
governed by {\bf TUT}. We return to this discussion elsewhere.\par

There will be some immediate results that the new resummed theory can
provide that amount to cross checks on it that are outside of the
regime of applicability of the superstring, for example. In the 
new approach, a point particle is something fundamental, in the 
superstring, a point particle is a large distance approximation.
Therefore, in the new theory there will be questions to answer that
do not occur in the superstring. One of these concerns the 
issue of the black hole character of a massive point particle
in the classical theory of general relativity. 
%%%%START HERE
%Specifically, as with any treatment of a classical system by
%quantum mechanical methods, our new approach to quantum
%gravity allows us to address 
%the interesting outstanding questions, 
%and there are many, posed by
%Einstein's theory
%to the quantum theory of point particle fields.
%Among these is the fate 
%of massive point particles that are so crucial to the success of
%the SM. 
For, in Einstein's theory, 
a point particle of non-zero rest mass $m$
has a non-zero Schwarzschild radius $r_S= 2(m/M_{Pl})(1/M_{Pl})$,
where $M_{Pl}$,~$1.22\times 10^{19}$ GeV, is the Planck mass,
so that such a particle should be a black hole~\cite{mtw} 
in the classical solutions
of Einstein's theory, unable to communicate ``freely'' with the world outside
of its Schwarzschild radius, except for some thermal effects
first pointed-out by Hawking~\cite{hawk}.
%%%START HERE 
Surely, this poses a problem for
the Standard Model phenomenology: it seems these point particles
are communicating freely their entire selves in their interactions with
each other. Can our new quantum theory of gravity reconcile
this apparent contradiction? It this question that we address here. 
\par

We start our analysis by setting up our new approach
to quantum gravity. As we explain in Ref.~\cite{bw1},
we follow the idea of Feynman~\cite{f1,f2} and treat
Einstein's theory as a point particle field theory
in which the metric of space-time undergoes quantum fluctuations
just like any other point particle does. On this view,
the Lagrangian density of the observable world is
\begin{equation}
%\begin{split}
{\cal L}(x) = -\frac{1}{2\kappa^2}\sqrt{-g} R
            + \sqrt{-g} L^{\cal G}_{SM}(x)
\label{lgwrld}
\end{equation}
where $R$ is the curvature scalar, $-g$ is the
negative of the determinant of the metric of space-time
$g_{\mu\nu}$, $\kappa=\sqrt{8\pi G_N}\equiv 
\sqrt{8\pi/M_{Pl}^2}$, where $G_N$ is Newton's constant,
and the SM Lagrangian density, which is well-known
( see for example, Ref.~\cite{sm,barpass} ) when invariance 
under local Poincare symmetry is not required,
is here represented by $L^{\cal G}_{SM}(x)$ which is readily obtained
from the familiar SM Lagrangian density as described in
Ref.~\cite{bw2}.
It is well-known that there are many massive 
point particles in (\ref{lgwrld}).
According to classical general relativity, they should all be black holes,
as we noted above. Are they black holes in our new approach to quantum gravity?
To study this question, we continue to follow Feynman in Ref.~\cite{f1,f2}
and treat spin as an inessential complication~\cite{mlg}, 
as the question of whether
a point particle with mass is or is not a black hole should not depend
too severely on whether or not it is spinning. We can come back to a 
spin-dependent analysis elsewhere~\cite{elswh}.\par 

Thus, we replace $L^{\cal G}_{SM}(x)$ in (\ref{lgwrld})
with the simplest case for our question, that of a free scalar field
, a free physical Higgs field, $\varphi(x)$, with a rest mass believed~\cite{lewwg} to be less than $400$ GeV and known to be greater than $114.4$ GeV with a
95\% CL. We are then led to consider the representative model
\begin{equation}
\begin{split}
{\cal L}(x) &= -\frac{1}{2\kappa^2} R \sqrt{-g}
            + \frac{1}{2}\left(g^{\mu\nu}\partial_\mu\varphi\partial_\nu\varphi - m_o^2\varphi^2\right)\sqrt{-g}\\
            &= \quad \frac{1}{2}\left\{ h^{\mu\nu,\lambda}\bar h_{\mu\nu,\lambda} - 2\eta^{\mu\mu'}\eta^{\lambda\lambda'}\bar{h}_{\mu_\lambda,\lambda'}\eta^{\sigma\sigma'}\bar{h}_{\mu'\sigma,\sigma'} \right\}\\
            & \qquad + \frac{1}{2}\left\{\varphi_{,\mu}\varphi^{,\mu}-m_o^2\varphi^2 \right\} -\kappa {h}^{\mu\nu}\left[\overline{\varphi_{,\mu}\varphi_{,\nu}}+\frac{1}{2}m_o^2\varphi^2\eta_{\mu\nu}\right]\\
            & \quad - \kappa^2 \left[ \frac{1}{2}h_{\lambda\rho}\bar{h}^{\rho\lambda}\left( \varphi_{,\mu}\varphi^{,\mu} - m_o^2\varphi^2 \right) - 2\eta_{\rho\rho'}h^{\mu\rho}\bar{h}^{\rho'\nu}\varphi_{,\mu}\varphi_{,\nu}\right] + \cdots \\
\end{split}
\label{eq1}
\end{equation}
Here, 
%$\varphi(x)$ is our representative scalar field for matter,
$\varphi(x)_{,\mu}\equiv \partial_\mu\varphi(x)$,
and $g_{\mu\nu}(x)=\eta_{\mu\nu}+2\kappa h_{\mu\nu}(x)$ 
where we follow Feynman and expand about Minkowski space
so that $\eta_{\mu\nu}=diag\{1,-1,-1,-1\}$. 
Following Feynman, we have introduced the notation
$\bar y_{\mu\nu}\equiv \frac{1}{2}\left(y_{\mu\nu}+y_{\nu\mu}-\eta_{\mu\nu}{y_\rho}^\rho\right)$ for any tensor $y_{\mu\nu}$\footnote{Our conventions for raising and lowering indices in the 
second line of (\ref{eq1}) are the same as those
in Ref.~\cite{f2}.}. 
Thus, $m_o$ is the bare mass of our free Higgs field and we set the small
tentatively observed~\cite{cosm1} value of the cosmological constant
to zero so that our quantum graviton has zero rest mass.
%Here, our normalizations are such that $\kappa=\sqrt{8\pi G_N}$
%where $G_N$ is Newton's constant.
The Feynman rules for (\ref{eq1}) have been essentially worked out by 
Feynman~\cite{f1,f2}, including the rule for the famous
Feynman-Faddeev-Popov~\cite{f1,ffp1} ghost contribution that must be added to
it to achieve a unitary theory with the fixing of the gauge
( we use the gauge of Feynman in Ref.~\cite{f1}, 
$\partial^\mu \bar h_{\nu\mu}=0$ ), 
so we do not repeat this 
material here. We turn instead directly to the issue 
of the effect of quantum loop corrections
on the black hole character of our massive Higgs field.
\par

\begin{figure}
\begin{center}
\epsfig{file=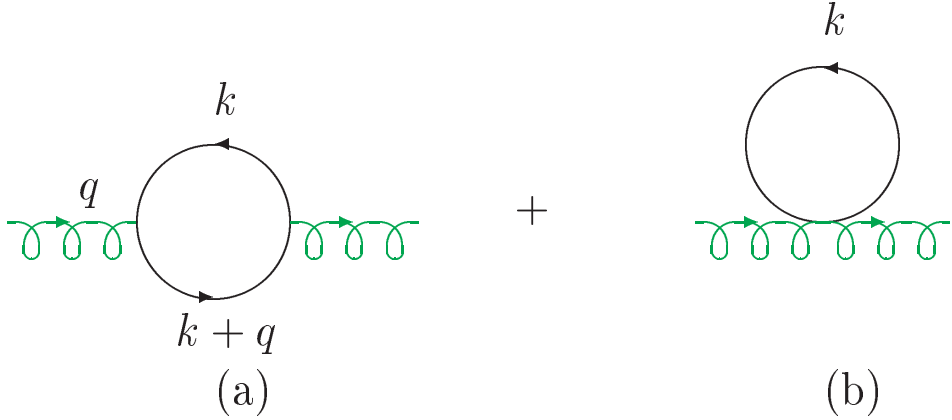,width=140mm}
\end{center}
\caption{\baselineskip=7mm     The scalar one-loop contribution to the
graviton propagator. $q$ is the 4-momentum of the graviton.}
\label{fig1}
\end{figure}

We initiate our approach by calculating the effects of the diagrams
in Fig.~\ref{fig1} on the graviton propagator. These effects give the 
possible one-loop corrections to
Newton's law that would follow from the matter in (\ref{eq1})
and will directly impact our black hole issue.
%It is sufficient to calculate the effects of the diagrams
%in Fig.~\ref{fig1} on the graviton 
%propagator to see the first quantum loop effect.
\par

In Ref.~\cite{bw1}, we have shown that, while
the naive power counting of
the graphs gives their degree of divergence as +4,
YFS~\cite{yfs} resummation of the soft graviton effects in the propagators
in Fig.~\ref{fig1} renders the graphs ultra-violet (UV) finite. 
Indeed, for example, for
Fig. 1a, we get without YFS resummation the result 
\begin{equation}
i\Sigma(q)^{1a}_{\bar\mu\bar\nu;\mu\nu}=\kappa^2\frac{\int d^4k}{2(2\pi)^4}
\frac{\left(k'_{\bar\mu}k_{\bar\nu}+k'_{\bar\nu}k_{\bar\mu}\right)
\left(k'_{\mu}k_{\nu}+k'_{\nu}k_{\mu}\right)}
{\left({k'}^2-m^2+i\epsilon\right)\left(k^2-m^2+i\epsilon\right)}
\label{eq2}
\end{equation},
where we set $k'=k+q$ and we take for definiteness only
fully transverse, traceless polarization states of the graviton to
be act on $\Sigma$ so that we have dropped the traces from its
vertices. Clearly, (\ref{eq2})
has degree of divergence +4. When we take into 
account the resummation as calculated in Ref.~\cite{bw1},
the free scalar propagators are improved to their YFS-resummed values,
\begin{equation}
i\Delta'_F(k)|_{\text{resummed}} =  \frac{ie^{B''_g(k)}}{(k^2-m^2+i\epsilon)}
\label{resum}
\end{equation},
where the virtual graviton function $B''_g(k)$ is, for Euclidean momenta,
\begin{equation}
B''_g(k) = \frac{\kappa^2|k^2|}{8\pi^2}\ln\left(\frac{m^2}{m^2+|k^2|}\right),
\label{deep}
\end{equation}  
so that we get instead of (\ref{eq2}) the result
( here, $k\rightarrow (ik^0,\vec k)$ by Wick rotation )
\begin{equation}
i\Sigma(q)^{1a}_{\bar\mu\bar\nu;\mu\nu}=i\kappa^2\frac{\int d^4k}{2(2\pi)^4}
\frac{\left(k'_{\bar\mu}k_{\bar\nu}+k'_{\bar\nu}k_{\bar\mu}\right)e^{\frac{\kappa^2|{k'}^2|}{8\pi^2}\ln\left(\frac{m^2}{m^2+|{k'}^2|}\right)}
\left(k'_{\mu}k_{\nu}+k'_{\nu}k_{\mu}\right)e^{\frac{\kappa^2|k^2|}{8\pi^2}\ln\left(\frac{m^2}{m^2+|k^2|}\right)}}
{\left({k'}^2-m^2+i\epsilon\right)\left(k^2-m^2+i\epsilon\right)}.
\label{eq2p}
\end{equation}
Evidently, this integral converges; so does that for Fig.1b when
we use the improved resummed propagators. This means that we
have a rigorous quantum loop correction to Newton's law
from Fig.1 which is finite and well defined.\par

What is the true difference between our new approach and just introducing 
a cut-off of the form suggested by our resummed propagators by hand
into the Feynman series? The difference is in the respective YFS residuals
defined in Ref.~\cite{bw1} and the references therein. 
In our new theory, these residuals are defined in such a way that
the resummed series is exactly equal to the original Feynman series.
In a by-hand cut-off, variants of which are often 
employed in the renormalization program
for a renormalizable theory for example, the series with the cut-off only approaches the original series as the cut-off is removed: as long as the cut-off
is present, the two series {\it are not equal}. We can not stress this
difference too much.\par

To see how our new result for Fig.~\ref{fig1} 
impacts the black hole character of our
massive point particle, we continue to work in the transverse, traceless
space for the graviton self-energy $\Sigma$\footnote{ As all physical polarization states are propagated with the same Feynman denominator, any physical
subspace can be used to determine this denominator. } and  
we get, to leading order, that the graviton propagator denominator
becomes
\begin{equation}
q^2 +\frac{1}{2}q^4\Sigma^{T(2)}+i\epsilon
\label{prop}
\end{equation}
where the transverse, traceless self-energy function $\Sigma^T(q^2)$ follows from eq.(\ref{eq2p}) for Fig. 1a and its analog for Fig. 1b
by the standard methods. 
For the coefficient of $q^4$ in $\Sigma^T(q^2)$ for $|q^2|>>m^2$
we have the result
\begin{equation} 
-\frac{1}{2}\Sigma^{T(2)} \cong \frac{c_2}{360\pi M_{Pl}^2}
\label{sigma}
\end{equation}
for
\begin{equation} 
c_2 = \int^{\infty}_{0}dx x^3(1+x)^{-4-\lambda_c x}\cong 72.1 
\label{int1}
\end{equation}
where $\lambda_c=\frac{2m^2}{\pi M_{Pl}^2}$.
When we Fourier transform the
inverse of (\ref{prop}) we find the potential
\begin{equation}
\Phi_{Newton}(r)= -\frac{G_N M_1M_2}{r}(1-e^{-ar})
\end{equation}
where $a=1/\sqrt{-\frac{1}{2}\Sigma^{T(2)}}\simeq 3.96 M_{Pl}$
in an obvious notation, where for definiteness, we set $m\cong 120$GeV.\par

At this point, let us note that the integral in (\ref{int1}) can be represented
for our purposes by the analytic expression~\cite{elswh}
\begin{equation}
c_2 \cong \ln\frac{1}{\lambda_c}-\ln\ln\frac{1}{\lambda_c}-\frac{\ln\ln\frac{1}{\lambda_c}}{\ln\frac{1}{\lambda_c}-\ln\ln\frac{1}{\lambda_c}}-\frac{11}{6}
\label{anal1}
\end{equation}
and we used this result to check the numerical result given in (\ref{int1}).
It is clear that, without resummation, we would have
$\lambda_c=0$ and our result in (\ref{int1})
would be infinite and, since this is the coefficient of $q^4$ in 
the inverse propagator,
no renormalization of the field and of the mass could be used to remove
such an infinity. In our new approach to quantum gravity, this 
infinity is absent.\par

In Ref.~\cite{bw2}, we have made a check on the gauge invariance 
of our result by comparing with the pioneering 
gauge invariant analysis of Ref.~\cite{thvelt1},
where the complete results for the one-loop divergences of our scalar
field coupled to Einstein's gravity have been computed. As we show in Ref.~\cite{bw2}, when the proper mapping of our $\lambda_c$ into the dimensional regularization parameter $2-n/2$ is done, we get complete agreement between
our result in  (\ref{sigma}) and the 
results in Ref.~\cite{thvelt1}. Here, $n$ is the analytically
continued dimension of space-time in the dimensional regularization
scheme~\cite{thvelt} of 't Hooft and Veltman and is understood
to be considered as $n\rightarrow 4$.
%We stress that our result in (\ref{sigma}) is gauge invariant, as our 
%approach involves the exact rearrangement of the Feynman series
%as we explain in Ref.~\cite{bw1} and the original series is gauge
%invariant. Indeed, one can cross check this result by comparing with
%the pioneering work in Ref.~\cite{thvelt1}, where the complete
%result of the one-loop divergences of our scalar field coupled
%to Einstein's gravity have been computed. This is made possible by the
%following observation. As we just observed, the result which we
%have obtained would be UV divergent without our resummation. Thus,
%the dominant terms which we are isolating in this paper
%are precisely those that
%are given in Ref.~\cite{thvelt1}, where we need to make the correspondence
%between the poles in $n$, the dimension of space-time,
%at $n=4$ calculated in Ref.~\cite{thvelt1} and the leading 
%log $\ln\frac{1}{\lambda_c}$. This we do by setting 
%the result $c_2$ equal to its value when $\lambda_c = 0$ in $n$ dimensions
%and allowing $n \rightarrow 4$. In this way we find that
%\begin{equation}
%1/(2-n/2) \leftrightarrow c_2 .
%\label{crsp1}
%\end{equation}
%This means that, if we look at the limit $q^2\rightarrow 0$,
%we get the result that the coefficient of $q^4$ 
%in  (\ref{prop}) is $3/(2-n/2)$ times the coefficient 
%of $c_2$ on the right-hand side of (\ref{sigma}), and this is in
%complete agreement with the result that is implied by
%eq.(3.40) in  Ref.~\cite{thvelt1}, for example. Of course, the results in
%Ref.~\cite{thvelt1} are also gauge invariant.\par

In the SM, there are now believed to be three massive neutrinos~\cite{neut},
with masses that we estimate at $\sim 3$ eV, and the remaining members
of the known three generations of Dirac fermions $\{e,\mu,\tau,u,d,s,c,b,t\}$, 
with masses given by ~\cite{pdg2002},
$m_e\cong 0.51$ MeV, $m_\mu \cong 0.106$ GeV, $m_\tau \cong 1.78$ GeV,
$m_u \cong 5.1$ MeV, $m_d \cong 8.9$ MeV, $m_s \cong 0.17$ GeV,
$m_c \cong 1.3$ GeV, $m_b \cong 4.5$ GeV and $m_t \cong 174$ GeV,
as well as the massive vector bosons $W^{\pm},~Z$, with masses 
$M_W\cong 80.4$ GeV,~$M_Z\cong 91.19$ GeV. Using the general spin independence
of the graviton coupling to matter at {\it relatively low} momentum transfers,
we see that we can take the effects of these degrees of freedom into account
approximately by counting each Dirac fermion as 4 degrees of freedom,
each massive vector boson as 3 degrees of freedom and remembering that
each quark has three colors. Using the result (\ref{anal1}) for each
of the massive degrees of freedom in the SM, we see that
the effective value of $c_2$ in the SM is approximately
\begin{equation}
c_{2,eff} \cong 9.26\times 10^3
\label{ceff} 
\end{equation}
so that the effective value of $a$ in the SM is 
\begin{equation}
a_{eff} \cong 0.349 M_{Pl} .
\label{aeff} 
\end{equation}
To make direct contact with black hole physics, note that,
for $r\rightarrow r_S$, $a_{eff}r \ll 1$ so 
that $|2\Phi_{Newton}(r)|_{M_1=m}/M_2|\ll 1$.
This means that in the respective solution for our metric of space-time,
$g_{00}\cong 1+2\Phi_{Newton}(r)|_{M_1=m}/M_2$ remains 
positive as we pass through the
Schwarzschild radius. Indeed, it can be shown that this 
positivity holds to $r=0$. Similarly, $g_{rr}$ remains negative
through $r_S$ down to $r=0$. To get these results,
note that in the relevant regime for r, the smallness of
the quantum corrected Newton potential means that we can use the
linearized Einstein equations for a small spherically symmetric
static source $\rho(r)$ which generates $\Phi_{Newton}(r)|_{M_1=m}/M_2$
via the standard Poisson's equation. 
The usual result~\cite{abs,mtw,sw1} for the
respective metric solution then gives 
$g_{00}\cong 1+2\Phi_{Newton}(r)|_{M_1=m}/M_2$ and
$g_{rr}\cong -1+2\Phi_{Newton}(r)|_{M_1=m}/M_2$ which remain
respectively time-like and space-like
to $r=0$.\par
It follows that the quantum corrections have obviated the classical
conclusion that a massive point particle is a black hole~\cite{mtw}.\par

The reader may wonder rightfully 
about the connection, if any, between
our resummed theory and the asymptotic safety approach, as in both cases
there is improved UV behavior. The similarity is even deeper because
the results in eq.(7)-eq.(10) and eq.(13) imply that
we can interpret our result for the correction to Newton's law as
a running Newton's constant 
$$G_N(k)=G_N/(1+\frac{k^2}{a_{eff}^2})\qquad \qquad $$
and this implies fixed point behavior for $k^2\rightarrow \infty$.
Indeed, even the value we get for $a_{eff}$ is similar to that
found in Ref.~\cite{reuter2} where an explicit realization of the 
asymptotic safety approach is presented. One can see that our approach
gives another explicit realization of asymptotic safety which however
does not involve any unknown parameters or arbitrary cut-off 
functions.
\par

Indeed, the result that a point particle of the SM with
non-zero rest mass does not have
a horizon is similar to the result in Ref.~\cite{reuter2},
derived using the asymptotic safety approach, that a black hole
with mass less than a critical mass $M_{cr}\sim M_{Pl}$ does not have
a horizon. The basic mechanism that drives the two results is the same, the
effective value of Newton's constant for the sub-Planck scale regime is very
weak because $G_N(k)$ vanishes for $k^2\rightarrow \infty$.\par

The agreement between our approach and the results in Ref.~\cite{reuter2}
goes further. In Ref.~\cite{reuter2}, the final state of the Hawking radiation
of an originally very massive black hole was found to be 
a Planck scale remnant. Here, if we use the results
in Ref.~\cite{reuter2} for the connection between $k$ and $r$
for the regime in which the lapse function vanishes,
then the similarity in our values for the coefficient of
$k^2$ in the denominator of $G_N(k)$ leads us exactly to 
the same Hawking radiation phenomenology for massive black holes
as was found in Ref.~\cite{reuter2}:
as the black hole evaporates, it reaches a critical mass
$M_{cr}\sim M_{Pl}$ at which the Bekenstein-Hawking temperature
vanishes, leaving a Planck scale remnant as the final state
of the Hawking evaporation process. 
\par

We do not wish to suggest that the value of $a_{eff}$ given here is complete,
as there may be as yet unknown massive particles beyond those already
discovered. Including more particles in the computation of
$a_{eff}$ would make it smaller and hence would not change the
conclusions of our analysis. For example, in the Minimal 
Supersymmetric Standard
Model we expect approximately that 
$a_{eff}\rightarrow \frac{1}{\sqrt{2}}a_{eff}$.
In addition, we point-out that, using the appropriate mapping
between $\lambda_c$ and $2-n/2$ as discussed above,
%correspondence in (\ref{crsp1}) 
one can also use the results for the
complete one-loop corrections in Ref.~\cite{thvelt1}
to the theory treated here to see that the remaining 
interactions at one-loop order not discussed here 
(vertex corrections, pure gravity self-energy corrections, etc. )
also do not increase the value of $a_{eff}$~\cite{elswh} ( This
follows from the use of the equations of motion and the comparison
of the coefficient of $R_{\mu\nu}^2$ in eqs.(5.22) and (5.23)
in Ref.~\cite{thvelt1}. ).
We can thus think of $a_{eff}$
as a parameter which is bounded from above by the estimates we give
above and which should be determined from cosmological and/or other
considerations. Further such implications will be taken up elsewhere.\par

\section*{Acknowledgements}

We thank Profs. S. Bethke and L. Stodolsky for the support and kind
hospitality of the MPI, Munich, while a part of this work was
completed. We thank Prof. C. Prescott for
the kind hospitality of SLAC Group A while this
work was in progress. We thank Prof. S. Jadach for useful discussions.

\end{document}